\documentclass[3p,times,sort&compress]{elsarticle}
\usepackage{ecrc}
\usepackage{amssymb}
\usepackage{amsmath}
\usepackage[figuresright]{rotating}
\usepackage[utf8]{inputenc}
\volume{00}

\firstpage{1}
\journalname{Physics Letters A}
\runauth{S.M. Kopeikin, E.M. Mazurova, A.P. Karpik}
\jnltitlelogo{Physics Letters A}
\jid{PLA}
\CopyrightLine{2011}{Published by Elsevier Ltd.}
\usepackage{bm}
\usepackage{url}
\usepackage{color}
\expandafter\ifx\csname package@font\endcsname\relax\else
 \expandafter\expandafter
 \expandafter\usepackage
 \expandafter\expandafter
 \expandafter{\csname package@font\endcsname}%
\fi

\def\be{\begin{equation}}
\def\ee{\end{equation}}
\def\ba{\begin{eqnarray}}
\def\ea{\end{eqnarray}}

\def\a{\alpha}
\def\a {\alpha}
\def\b{\beta}
\def\g{\gamma}     
     \def\D{\Delta}

\def\e{\epsilon}

\def\m{\mu}
\def\n{\nu}
\def\o{\omega}   \def\O{\Omega}

\def\th{\theta}

\def\r{\rho}

\def\t{\tau}

\def\la{\label}
\def\pd{\partial}
\def\lt{\left}
\def\ri{\right}

\begin{document}
\begin{frontmatter}
\title{Towards exact relativistic theory of Earth's geoid undulation}
\author[av1,av3]{Sergei M. Kopeikin\corref{cor1}}
\ead{kopeikins@missouri.edu}
\cortext[cor1]{Corresponding author}
\author[av2,av3]{Elena M. Mazurova}
\ead{e\_mazurova@mail.ru}
\author[av3]{Alexander P. Karpik}
\ead{rector@ssga.ru}

\address[av1]{Department of Physics \& Astronomy, University of Missouri, Columbia, Missouri 65211, USA}
\address[av2]{Moscow State University of Geodesy and Cartography, 4 Gorokhovsky Alley, Moscow 105064, Russia}
\address[av3]{Siberian State Geodetic Academy, 10 Plakhotny St., Novosibirsk 630108, Russia}

\begin{abstract}
The present paper extends the Newtonian concept of the geoid in classic geodesy towards the realm of general relativity by utilizing the covariant geometric methods of the perturbation theory of curved manifolds. It yields a covariant definition of the anomalous (disturbing) gravity potential and formulate differential equation for it in the form of a covariant Laplace equation. The paper also derives the Bruns equation for calculation of geoid's height with full account for relativistic effects beyond the Newtonian approximation. A brief discussion of the relativistic Bruns formula is provided.
\end{abstract}
\begin{keyword}
gravity \sep relativity \sep geodesy \sep geoid \sep undulation  
\PACS 04.20.-q \sep 04.25.Nx \sep 91.10.-v \sep 91.10.By 
\end{keyword}

\end{frontmatter}
\section{Introduction} 
\noindent

Knowledge of the figure and size of the Earth is vitally important in geophysics and in applied sciences for determining precise position of objects on Earth's surface and in near space, depicting correctly topographic maps, creating digital terrain models, and many others. Solution of this problem is challenging for the real figure of the Earth has an irregular shape which can be neither described by a simple analytic expression nor easily computed as mass distribution of the Earth is not known well enough \citep{Moritz_1967}. To manage solution of this problem, C.~F. Gauss proposed to take one of the equipotential surfaces of Earth's gravitational field as a mathematical idealization approximating the real shape of the Earth such that it coincides with the mean sea level of idealized oceans representing the surface of homogeneous water masses at rest, subject only to the force of gravity and free from variations with time \citep{Torge_2012_book}.  In 1873, a German mathematician J.~B. Listing \footnote{It is the same J.~B. Listing who introduced in 1847 the term {\it topology} in mathematics.} coined the term {\it geoid} to describe this mathematical surface and, since then, the geoid has become a subject of a considerable scientific investigation in geodesy, oceanography, geophysics, and other Earth sciences \citep{Amalvict_geoid_survey}. Geoid's equipotential surface is perpendicular everywhere to the gravity force vector defining direction of the plumb line. In its own turn, the direction of  plumb line is defined by the law of distribution of mass density inside Earth's crust and mantle. For the mass distribution is basically uneven, the shape of geoid's surface is not an ellipsoid of revolution with regularly varying curvature. 

The Stokes-Poincar\'e theorem has played a major role in developing the theory of Earth's figure: if a body of total mass $M$ rotates with constant angular velocity $\O$ about a fixed axis, and if ${\cal S}$ is a level surface of its gravity field enclosing the entire mass, then the gravity potential in the exterior space of ${\cal S}$ is uniquely determined \footnote{In classic geodesy Earth's angular velocity is denoted $\o$. However, this symbol is commonly used in general relativity to denote vorticity, and we do not employ it in this paper.} by $M$, $\O$, and the parameters defining ${\cal S}$ \citep{Torge_2012_book}. However, geodesy is more interested in the inverse problem of the theory of Earth's figure which is to determine the shape of geoid from observed values of gravitational field. 

Geoid's precise calculation is usually carried out by combining a global geopotential model of gravitational field with terrestrial gravity anomalies measured in the region of interest and supplemented with the local/regional topographic information. The gravity anomalies (along with other modern methods \citep{Torge_2012_book}) allow us to find out the undulation of geoid's surface that is measured with respect to a reference level surface of the World Geodetic System \citep{Lohmar_1988} established in 1984 (WGS84), and last revised in 2004. This reference surface is called a reference ellipsoid. Geoid's undulation is given in terms of height above the ellipsoid taken along the normal line to ellipsoid's surface (see \textcolor{blue}{\url{http://earth-info.nga.mil/GandG/wgs84/}} for more detail).

A reference level surface, $\bar{\cal S}$, is defined by the condition of constant gravity potential, $\bar U_{\rm N}$, generated by a perfect fluid being rigidly rotated with respect to celestial reference frame \citep{Soffel_2003AJ} with a constant angular velocity $\O$,
\be\la{1}
\bar U_{\rm N}(r,\th)\equiv \bar V(r,\th)+\frac12\Omega^2r^2\sin^2\th\;,
\ee
where $x^i=\{x^1,x^2,x^3\}=\{r,\th,\lambda\}$ are spherical coordinates: $r$ - radius-vector, $\th$ - the polar angle (co-latitude) measured from the rotational axis, and $\lambda$ - the longitude measured in the equatorial plane. Equation \eqref{1} also defines the surfaces of constant density and pressure of the fluid \citep{Torge_2012_book}. 

The quantity $\bar V=\bar V(r,\th)$ in \eqref{1} is the axisymmetric gravitational potential determined inside the mass distribution by the Poisson equation,
\be\la{2} 
\D_{\rm N}\bar V(r,\th)=-4\pi G\bar\rho\;,
\ee
where $\bar\rho=\bar\rho(r,\th)$ is the axisymmetric volume mass density, $G$ is the Newtonian gravitational constant, 
\be\la{1aa}
\D_{\rm N}\equiv\pd_{rr}+\frac{2}{r}\pd_r+\frac1{r^2}\pd_{\th\th}+\frac1{r^2\tan\th}\pd_\th+\frac1{r^2\sin^2\th}\pd_{\lambda\lambda}\;,
\ee
is the Laplace operator in spherical coordinates, and the partial derivatives $\pd_i\equiv\pd/\pd x^i$, $\pd_{ij}\equiv\pd^2/\pd x^i\pd x^j$ (the Roman indices take values $1,2,3$). Inside masses a differential equation defining the gravity potential, $\bar U_{\rm N}$, is
\be\la{pm8y}
\D_{\rm N}\bar U_{\rm N}= -4\pi G\bar\rho+2\O^2\;.
\ee
and is mostly used in geophysics. 

Physical geodesy uses the Laplace equation
\be\la{2df} 
\D_{\rm N}\bar V(r,\th)=0\;,
\ee
instead of \eqref{2} as the gravitational field is only required outside masses for all relevant applications. Laplace's equation \eqref{2df} is fully sufficient to determine the gravitational potential $V$ in the exterior space, where the density distribution has not to be known. Nonetheless, it is worth emphasizing that solution of the Laplace equation \eqref{2df} is not fully arbitrary but must match with a solution of the Poisson equation \eqref{2} with physically meaningful mass density distribution inside Earth's body.

Because all functions depend only on $r$ and $\th$, the reference surface is an axisymmetric body. In the most general case, equation \eqref{1} does not define a surface of the ellipsoid of revolution. Only in case of a uniform mass density, $\bar\rho={\rm const.}$, the reference level surface coincides with the ellipsoid of revolution \citep[section 5.2]{Tassoul_1978_book}. The homogeneous ellipsoid of revolution is very convenient as a reference surface because its external (called {\it normal}) gravity field can be modelled by closed formulas. In principle, it is possible to construct level spheroids that provide a better fit to the geoid but their equations are more complicated mathematically and do not significantly reduce deviation of geoid from  ellipsoid. Hence, they are less suitable as physical normal figures \citep[Section 4.2.1]{Torge_2012_book}. 

When applying general relativity to calculation of geoid's surface, it becomes important to realize that the post-Newtonian reference level surface cannot be the ellipsoid of revolution any longer. The reason is that a figure of reference in geodesy is to be a solution of the Newtonian gravity field equation (\ref{pm8y}). The same principle must be hold in general relativity. It requires to find out an exact interior solution of the Einstein gravity field equations that would be consistent with the solution representing the homogeneous ellipsoid of revolution in classic geodesy. This general-relativistic problem is not trivial from mathematical point of view, because of non-linearity of Einstein's equations, and has not yet been solved. Calculations conducted in the post-Newtonian approximations reveal that the uniformly rotating perfect fluid with homogeneous density is not an ellipsoid but represents an axisymmetric surface of a higher polynomial order \citep{Chandr_1967ApJ147,Bardeen_1971ApJ,Petroff_2003PhRvD,Meinel_2012} but the convergence of the post-Newtonian series has not yet been explored. In this situation, the only restriction which we impose in the present paper on the shape of the reference level surface is that it is consistent with the Einstein equations.

Earth's crust is a thin surface layer having irregular mass density that deviates significantly from the axisymmetric distribution. Furthermore, the Earth mantle shows a non-axisymmetric surface deformation which easily reaches the same dimension as the crust variation, and its density is much bigger than the density of the crust. Because of these irregularities in both crust and mantle, the physical surface, ${\cal S}$, of the geoid is perturbed and deviates from the equipotential surface $\bar{\cal S}$ of the unperturbed (axisymmetric) figure defined by \eqref{1}. We introduce the overall mass density perturbation of both the mantle and the crust by equation
\be\la{4qq}
\mu(r,\th,\lambda)\equiv \rho(r,\th,\lambda)-\bar\rho(r,\th)\;,
\ee
where $\rho(r,\th,\lambda)$ is the actual density of Earth's matter. We denote the actual gravity potential of Earth by
\be\la{3edg}
W_{\rm N}(r,\th,\lambda)\equiv V(r,\th,\lambda)+\frac12\Omega^2r^2\sin^2\th\;,
\ee
where $V=V(r,\th,\lambda)$ is a gravitational potential that is determined by the Poisson equation
\be\la{3}
\D_{\rm N} V(r,\th,\lambda)=-4\pi G\rho(r,\th,\lambda)\;,
\ee
inside masses, and the Laplace equation
\be\la{3opk}
\D_{\rm N} V(r,\th,\lambda)=0\;,
\ee
outside masses.

We call the difference 
\be\la{4}
T_{\rm N}(r,\th, \lambda)\equiv W_{\rm N}(r,\th, \lambda)-\bar U_{\rm N}(r,\th)\;,
\ee
the disturbing (Newtonian) potential where both functionals, $W_{\rm N}$ and $\bar U_{\rm N}$,  are calculated at the same point of space under assumption that the angular velocity $\O$ remains unperturbed. 
It is straightforward to see that the disturbing potential obeys to 
\be\la{5}
\D_{\rm N} T_{\rm N}(r,\th, \lambda)=-4\pi G\mu(r,\th, \lambda)\;,
\ee
inside the mass distribution, and to the Laplace equation
\be\la{6}
\D_{\rm N} T_{\rm N}(r,\th, \lambda)=0\;,
\ee
outside masses. 

Molodensky \citep{Molodenski_1958BGeod,Molodensky_1962BGeod} reformulated \eqref{6} into an integral equation 
\be\la{7}
2\pi T_{\rm N}+\oiint\limits_\Sigma\frac{T_{\rm N}}{\ell}n^i\pd_i\ln\left(\ell T_{\rm N}\ri)d\Sigma=0\;,
\ee 
where $\ell=|{\bm x}-{\bm x}'|$ denotes the distance between the source point, ${\bm x}'$, taken on Earth's surface $\Sigma$ and the field point, ${\bm x}$, while $d\Sigma$ is the surface element of integration at point ${\bm x}'$, and $n^i$ is the (outward) unit normal to $\Sigma$ at ${\bm x}'$. The physical surface $\Sigma$ of the Earth is known from the Global Navigation Satellite System (GNSS) measurements \citep{Moritz_1967}. Thus, the only remaining unknown in \eqref{7} is the disturbing potential, $T_{\rm N}$. It can be found from \eqref{7} by employing the gravity disturbances of $T_{\rm N}(\Sigma)$ taken on $\Sigma$ as boundary values \citep{Mazurova_2011}. As soon as $T_{\rm N}$ is known everywhere in space, the geoid's undulation (its height ${\mathfrak N}$ above the reference ellipsoid) can be found from Bruns' equation \citep{Moritz_1967}
\be\la{8}
{\mathfrak N}=\frac{T_{\rm N}({\cal S})}{\g_{\rm N}}\;,
\ee
where the potential $T_{\rm N}({\cal S})$ refers to geoid, and $\g_{\rm N}$ is the normal gravity on the reference ellipsoid (surface $\bar{\cal S}$). 

Producing a precise global map of the geoid's undulation has proven to be a challenge. The important discoveries in the classic (Stokes and Molodensky) theory of the geoid computation were made in XX-th century by a number of researchers (see review in \citep{Torge_2012_book}). The precision of geoid's computation on the global scale has been further improved in XXI-st century with the advent of  gravimetric satellites like GRACE (\textcolor{blue}{\url{http://www.csr.utexas.edu/grace}}) and GOCE (\textcolor{blue}{\url{http://www.esa.int/Our_Activities/Observing_the_Earth/GOCE}}). It will continue to improve as new geodetic data will be accumulating.  

General relativistic corrections to the Newtonian theory of geoid can reach magnitude of a centimetre \citep{Kopeikin_2011_book,Mai_2014}. Though this number looks small but it is within the range of modern geodetic techniques which now include, besides conventional sensors, also atomic clocks \citep{Mai_2013ZGGL,Falke_2013,Petit_2014} that allow us to measure the potential difference of gravitational field between two points directly instead of deducing it from the combination of geometric levelling and gravimetry. This is because the rate of clocks is fully determined by the metric tensor of relativistic theory of gravity. Therefore, taking into account relativistic corrections in the determination of geoid's undulation is getting practically important. Furthermore, there is a growing demand among geodetic community for merging the science of geodesy with modern theoretical description of space, time and gravity - the Einstein general relativity. It requires working out an exact relativistic theory of geodetic measurements. 

This paper extends the Newtonian theory of Earth's geoid and its undulation into the realm of general relativity. It is organized as follows. Section \ref{sec2} defines the background (axisymmetric) spacetime manifold and derives Einstein's equations for the unperturbed metric tensor. Section \ref{sec2aa} describes reference level surface. Section \ref{sec3} gives two definitions of the relativistic geoid and discusses their equivalence. Section \ref{sec4} introduces the general-relativistic, anomalous gravity potential. Section \ref{sec5} derives the master equation for the anomalous gravity potential. Finally, section \ref{sec6} yields the relativistic Bruns equation for geoid's undulation. 

We denote the speed of light $c$ and use the Einsteinian gravitational constant $\kappa=8\pi G/c^2$. Other notations are explained in the main text as they appear. 
\section{Background spacetime manifold}\la{sec2}
\noindent
Formulation of relativistic theory of Earth's geoid begins from the construction of an unperturbed (background) spacetime manifold associated with a uniformly rotating Earth under assumption that the tidal forces are neglected and Earth's matter has a stationary, axisymmetric distribution.  We use spherical coordinates $x^\a=\{x^0,x^1,x^2,x^3\}\equiv\{ct,r,\th,\lambda\}$ co-rotating with Earth rigidly with constant angular velocity, $\O$, counter-clockwise. The metric $\bar g_{\a\b}$ of the background manifold is defined as follows \citep{Gourgoulhon_2010arXiv}
\ba\la{9}
d\bar s^2&=&\bar g_{\a\b}dx^\a dx^\b\\\nonumber
&=&-\left[c^2N^2-(\O-\mathfrak{G})^2B^2r^2\sin^2\th\ri]dt^2+2(\O-\mathfrak{G})B^2 r^2\sin^2\th dtd\lambda+A^2\left(dr^2+r^2d\th^2\ri)+B^2 r^2\sin^2\th d\lambda^2\;,
\ea
where $N\equiv N(r,\th)$, $A\equiv A(r,\th)$, $B\equiv B(r,\th)$, $\mathfrak{G}\equiv\mathfrak{G}(r,\th)$ are functions of only two coordinates, $r$ and $\th$, and the Greek (spacetime) indices take values $0,1,2,3$, here and everywhere else \footnote{The lapse function $N$ should not be confused with geoid's height ${\mathfrak N}$ introduced earlier in \eqref{8}.}. The metric, $\bar g_{\a\b}$, and its inverse, $\bar g^{\a\b}$, are used for rising and lowering the Greek  indices. The repeated Greek indices denote the Einstein summation rule. 

We notice that the stationary, axisymmetric metric \eqref{9} possesses two Killing vectors corresponding to translations along time, $x^0\equiv ct$, and azimuthal, $x^3\equiv\lambda$, coordinates. In the Newtonian limit functions $A=B=1$, $\mathfrak{G}=0$, and $N=1-2\bar V/c^2$, where $\bar V$ is the Newtonian gravitational potential defined by equation \eqref{2}. General relativity predicts deviation of these functions from their Newtonian values. In particular, function $\mathfrak{G}$ represents a new type of gravitational field not being present in the Newtonian theory -- gravitomagnetic field -- that arises in general relativity due to the rotation of the Earth \citep{Ciufolini_1995book}. It is very weak but can be presently measured with satellite laser ranging technique \citep{Ciufolini_2012EPJP} and/or by means of a spinning gyroscope flying around Earth in a drag-free satellite \citep{Everitt_2011PhRvL}.  

Unperturbed four-velocity of Earth's matter, $\bar u^\a=c^{-1}dx^\a/d\bar\t$, where $\bar\t$ is the proper time taken along the world line of the mass element, $c^2d\bar\t^2=-d\bar s^2$. For the matter is at rest in the rotating coordinates, its four-velocity has the following components, $\bar u^\a=\left\{\bar u^0,u^r,u^\th,u^\lambda\ri\}=\left\{\bar u^0,0,0,0\ri\}$ where 
\be\la{10}
\bar u^0=\left[N^2-c^{-2}(\O-\mathfrak{G})^2B^2r^2\sin^2\th\ri]^{-1/2}\;.
\ee
World lines of the mass elements form a rotating and accelerating congruence without divergence. Indeed, the chronometric decomposition \citep{Zelmanov_2006_book} of a covariant derivative of four-velocity reads \citep{Novikov_1989_book} 
\be\la{10a}
\bar u_{\a|\b}=\bar \o_{\a\b}+\bar \sigma_{\a\b}+\frac13\bar\th\bar h_{\a\b}-\bar a_\a \bar u_\b\;,
\ee
where here, and everywhere else, the vertical bar denotes a covariant derivative on the background manifold with metric \eqref{9}. The quantity 
\be\la{10b}
\bar h_{\a\b}\equiv \bar g_{\a\b}+\bar u_\a\bar u_\b\;,
\ee
represents metric tensor on 3-dimensional hypersurfaces (spatial slices) being orthogonal to $\bar u^\a$, $\bar a^\a\equiv \bar u^\b\bar u_{\a|\b}$ is  four-acceleration, $\bar\th\equiv \bar u^\a{}_{|\a}$ -- divergence of the congruence (which should not be confused with spherical coordinate $\th$), and $\bar \sigma_{\a\b}$ and $\bar \o_{\a\b}$ are tensors of shear (deformation) and vorticity (rotation) of the congruence,
\ba\la{10c}
\bar \sigma_{\a\b}&\equiv&\frac12\lt(\bar u_{\a|\m}\bar h^\m{}_\b+\bar u_{\b|\m}\bar h^\m{}_\a\ri)-\frac13\bar\th\bar h_{\a\b}\;,\\\la{10d}
\bar \o_{\a\b}&\equiv&\frac12\lt(\bar u_{\a|\m}\bar h^\m{}_\b-\bar u_{\b|\m}\bar h^\m{}_\a\ri)\;.
\ea
In case of a rigidly rotating axisymmetric configuration we have $\bar\sigma_{\a\b}=\bar\th=0$ but $\bar a_\a\not=0$ because the matter particles do not move along geodesics, and $\bar \o_{\a\b}\not=0$ because the matter is rotating. Spatial metric \eqref{10b} is used to measure the proper (physical) distances in space \citep{Zelmanov_2006_book,LL_book2}. 

The symmetric energy-momentum tensor of the rotating matter
\be\la{11}
c^2\bar T^{\a\b}=\bar\r(c^2+\bar\Pi)\bar u^\a\bar u^\b+\bar p\bar h^{\a\b}+\bar\pi^{\a\b}\;,
\ee
where $\bar\r$ is the mass density, $\bar p$ -- pressure, $\bar\Pi$ -- the compression energy of matter, and $\bar\pi_{\a\b}$ is the tensor of residual stresses ($\bar\pi^{\a\b}\bar u_\a=0$). Pressure, density and the compression energy are related by the equation of state and by the thermodynamic laws. 

Einstein's field equations outside masses are
\be\la{enq1}
\bar R_{\a\b}=0\;,
\ee
and inside matter,
\be\la{21}
\bar R_{\a\b}=\kappa\lt(\bar T_{\a\b}-\frac12\bar g_{\a\b}\bar T\ri)\;,
\ee
where $\bar T\equiv \bar T^\a{}_\a=\bar g^{a\b}\bar T_{\a\b}$, $\bar R_{\a\b}$ is the Ricci tensor formed from metric tensor \eqref{9}, its first and second derivatives \citep[Section 3.7]{Kopeikin_2011_book}. In what follows, we operate with equation \eqref{enq1} which is equivalent to the Laplace equation in classic geodesy.

Substituting metric \eqref{9} and tensor (\ref{11}) to \eqref{enq1} or \eqref{21} yields differential equations for the four functions entering the metric. More practical for geodesy are the Einstein equations \eqref{enq1} in vacuum. In this case the Einstein equations \eqref{enq1} read \citep{Gourgoulhon_2010arXiv}
\ba
\la{13}\displaystyle
\left(\pd_{rr}+\frac{1}{r}\pd_r+\frac1{r^2}\pd_{\th\th}\ri)\lt(\ln A+\n\ri)&=&\frac{3B^2r^2\sin^2\th}{4c^2N^2}\pd\mathfrak{G}\pd\mathfrak{G}-\pd\n\pd\n\;,
\\\la{14}
\displaystyle
\left(\pd_{rr}+\frac{2}{r}\pd_r+\frac1{r^2}\pd_{\th\th}+\frac1{r^2\tan\th}\pd_\th\ri)\n&=&\frac{B^2r^2\sin^2\th}{2c^2N^2}\pd\mathfrak{G}\pd\mathfrak{G}-\pd\n\pd\left(\n+\ln B\right)\;,
\\
\la{15}\displaystyle
\left(\pd_{rr}+\frac{1}{r}\pd_r+\frac1{r^2}\pd_{\th\th}\ri)\lt(NB-1\ri)r\sin\th&=&0\;,
\\
\la{16}\displaystyle
\left(\pd_{rr}+\frac{2}{r}\pd_r+\frac1{r^2}\pd_{\th\th}+\frac1{r^2\tan\th}\pd_\th-\frac1{r^2\sin^2\th}\ri)\mathfrak{G}r\sin\th &=&r\sin\th\;\pd\mathfrak{G}\pd\left(\n-3\ln B\right)\;,
\ea
where $\n\equiv\ln N$, and we have used the following abbreviation \citep{Gourgoulhon_2010arXiv}  for the product of two arbitrary functions, $u$ and $w$,
\be\la{17}
\pd u\pd w\equiv (\pd_r u)(\pd_r w)+\frac1{r^2}(\pd_\th u)(\pd_\th w)\;.
\ee
After solving \eqref{13}-\eqref{16} we get a vacuum description of the background spacetime manifold in terms of functions $A,B,N,\mathfrak{G}$ entering the metric tensor \eqref{9} \footnote{The homogeneous equation \eqref{15} admits a choice of coordinates such that it has a simple solution $B(r,\theta)N(r,\theta)=1$. In this case, only three functions $A,N,\mathfrak{G}$ in metric \eqref{9} are sufficient to find the gravitational field in vacuum \citep{Stephani_2009_book}. Different choice of coordinates is offered by choosing the product $\lt(NB-1\ri)r\sin\th$ in the form of a harmonic polynomial in 2-dimensional space which satisfies equation \eqref{15}.}.

\section{Reference level surface}\la{sec2aa}
\noindent
Generalization of the reference ellipsoid of classic geodesy to relativity requires an exact, and asymptotically-flat solution of the Einstein equations \eqref{13}-\eqref{16} for the axisymmetric, stationary-rotating mass distribution. This problem is formidable as the Einstein equations are highly non-linear. Therefore, at the time being there are only a few known, exact exterior solutions of this type including the Tomimatsu-Sato and Kerr metrics but their extrapolation to the interior of a rotating extended body remains unknown \citep{Stephani_2009_book}. Exact interior solution that may correspond to some rotational matter configuration was found by Wahlquist \citep{Wahlquist_1968} but, unfortunately, extrapolation of Wahlquist's metric to the exterior space does not match the asymptotically-flat, Minkowsky metric, $\eta_{\a\b}$, at infinity \citep{Bradley_2000}.

Some progress has been made towards finding an approximate (post-Newtonian) interior solutions for the metric of a rigidly rotating perfect fluid \citep{Chandr_1965ApJ142,Bardeen_1971ApJ,Chandr_1974MNRAS167,Petroff_2003PhRvD}. These solutions are sufficient for practical applications in geodesy. Finding a shape of the {\it exact} reference level configuration in the  relativistic geodesy, if one exists, remains an open theoretical problem. Fortunately, a formal development of general-relativistic theory of Earth's geoid undulation only requires the existence of such a reference level surface. We shall adopt this assumption. 

In any case, the reference configuration must be bounded by an equipotential level surface, $\bar U\equiv \bar U(r,\th)={\rm const.}$, where the relativistic gravity potential $\bar U$ is defined by the derivative of the proper time $\bar\t$ of metric \eqref{9}, 
\be\la{levs1}
\bar U=c^2\left[1-\lt(\frac{d\bar\t}{dt}\ri)\right]_{r,\th,\lambda\;{\rm fixed}}\;,
\ee
that is equivalent to $\bar U/c^2=1-1/\bar u^0$ where $\bar u^0=dt/d\bar\t$ is the time component of four-velocity of Earth's matter measured on the equipotential surface. Equation \eqref{levs1} extends the concept of the Newtonian gravity potential, $\bar U_{\rm N}$ given in \eqref{1}, to relativity. After picking up the value of $\bar u^0$ from \eqref{10}, equation \eqref{levs1} becomes,
\be\la{19}
\bar U(r,\th)\equiv c^2\left[1-\sqrt{N^2-c^{-2}(\O-\mathfrak{G})^2B^2r^2\sin^2\th}\ri]\;.
\ee
In the Newtonian approximation $N(r,\th)= 1-2\bar V(r,\th)/c^2$, $B(r,\th)= 1$ and $\mathfrak{G}(r,\th)= 0$. Expanding the root square in (\ref{19}) into the post-Newtonian series yields $\bar U(r,\th)\simeq \bar U_{\rm N}(r,\th)+{\cal O}(c^{-2})$, that matches the Newtonian definition (\ref{1}).

Differential equation for the relativistic potential, $\bar U$, is derived from the Landau-Raychaudhuri equation \citep[p. 84]{Hawking_1973_book} applied to the world lines of the reference frame rigidly rotating along with Earth's matter. Tensor of deformation, $\bar\sigma_{\a\b}$, of such a frame vanishes identically and the Landau-Raychaudhuri equation takes on the following form \citep[Problem 14.10]{LPPT_1975_book}
\be\la{20}
\bar h^{\a\b} a_{\a|\b}=\bar R_{\a\b}\bar u^\a\bar u^\b-\bar a_\a \bar a^\a-2\bar\omega^2\;,
\ee
where $\bar\omega^2\equiv (1/2)\bar \o_{\a\b}\bar \o^{\a\b}$, and we notice that in the Newtonian approximation the magnitude of the vorticity, $\bar\omega^2\simeq\O^2/c^2$. 

Stationary axisymmetric spacetime admits two Killing vectors, $\xi^\a=\pd_t$ and $\chi^\a=\pd_\lambda$, associated with translations along $t$ and $\lambda$ coordinates respectively \citep{Wald_1984_book}. Existence of the Killing vectors allows us to represent the four-acceleration of the  congruence in the form of a gradient taken from the time component of the four-velocity, $\bar a_\a=-\pd_\a\ln\bar u^0$, where $\bar u^0=\left(-\bar g_{00}\ri)^{-1/2}=\left(-\xi_\a\xi^\a\ri)^{-1/2}$ is interpreted as a scalar \citep[Problem 10.14]{LPPT_1975_book}. After accounting for \eqref{19} it yields \footnote{Four-acceleration $\bar a_\a$ is orthogonal to four-velocity, $\bar u^\a \bar a_\a=0$, and hence, is a purely spatial vector. Its spatial components relate to the acceleration of gravity, $\bar \g^i$, measured by accelerometer (gravimeter) as follows, $\bar\gamma^i\equiv -c^2 \bar a^i$.}
\be\la{22}
\bar a_\a=\pd_\a\ln\left(1-\frac{\bar U}{c^2}\ri)\;.
\ee
Replacing (\ref{22}) in \eqref{20} brings about a highly non-linear equation for potential $\bar U$,
\be\la{23}
\Delta\bar U-2\lt(\bar\omega^2+\bar a_\a\bar a^\a\ri)\left(c^2-\bar U\ri)=-8\pi G\lt(\bar T_{\a\b}\bar u^\a\bar u^\b+\frac12\bar T\ri)\left(1-\frac{\bar U}{c^2}\ri)\;.
\ee
where $\bar a_\a$ is given in \eqref{22}, $\bar\omega^2$ is a function of $\bar U$ and $\Omega-\mathfrak{G}$, and 
\be\la{23a}
\Delta \bar U\equiv\bar h^{\a\b}(\bar h^\m{}_\a \bar U_{|\m})_{|\b}\;,
\ee
is the covariant form of the Laplace operator of the spatial metric \eqref{10b}. In the Newtonian limit, $\bar U\simeq \bar U_{\rm N}$, and relativistic equation (\ref{23}) is reduced to \eqref{1aa}. Effectively, equation \eqref{23} can be solved only in combination with the Einstein equation (\ref{16}) for function $\mathfrak{G}$. 


It is worth noticing that if the Earth's matter were a rigidly rotating perfect fluid its equipotential surface would coincide with a surface of fluid's constant pressure. Indeed,
relativistic Euler's equation for the perfect fluid is \citep[Problem 14.3]{LPPT_1975_book}
\be\la{25}
\lt(\bar\e+\bar p\ri)\bar a_\a=-\pd_\a \bar p-\bar u_\a\bar u^\b\pd_\b\bar p\;,
\ee
where $\bar\e\equiv\bar\rho\left(c^2+\bar\Pi\ri)$.
A second term in the right-hand side of this equation vanishes because in stationary case pressure, $\bar p$, does not depend on time and, hence, $u^\b\pd_\b\bar p=u^0\pd_0\bar p=0$.
Contracting \eqref{25} with an infinitesimal vector of displacement, $dx^\a$, yields
\be\la{26}
d\bar p=-\lt(\bar\e+\bar p\ri)d\ln\left(1-\frac{\bar U}{c^2}\ri)\;.
\ee
The right-hand side of \eqref{26} vanishes on the equipotential surface which means that pressure, $\bar p={\rm const.}$ It can be shown \citep[Problem 16.18]{LPPT_1975_book} that the density, $\bar\rho$, and the specific internal energy, $\bar\Pi$, are also constant on the level surfaces.  

\section{Relativistic geoid}\la{sec3} 
\noindent
Pioneering study of relativistic geodesy including the geoid definition have been carried out by Bjerhammar \citep{Bjerhammar_1985}.
The Newtonian concept of Earth's geoid was extended to the post-Newtonian approximation of general relativity in \citep{Soffel_1989_book,Kopejkin_1991}. More recent discussion of the post-Newtonian gravimetry and geodesy is given in \citep{Kopeikin_2011_book,Mai_2014}. In this section we make a next step and introduce an {\it exact} concept of the relativistic geoid in general relativity that is not limited to the first post-Newtonian approximation. 

In real physical situations the background spacetime manifold is perturbed because the actual mass distribution, stresses, and velocity flow of Earth's matter is not axisymmetric. The angular velocity, $\Omega$, of Earth's rotation also changes because of precession, nutation, polar motion and variations in the length-of-day.  
The perturbed physical metric, $g_{\a\b}\equiv g_{\a\b}(t,r,\th,\lambda)$, depends on time and all three spatial coordinates, and can be split into an algebraic sum of the background metric \eqref{9}, and its perturbation, $\varkappa_{\a\b}\equiv\varkappa_{\a\b}(t,r,\th,\lambda)$, as follows
\be\la{27}
g_{\a\b}=\bar g_{\a\b}+\varkappa_{\a\b}\;.
\ee 
In the present paper, we shall neglect dependence of perturbation $\varkappa_{\a\b}$ on time because it produces very tiny relativistic effects that are currently unobservable.

Terrestrial reference frame is formed by the world lines of observers having fixed spatial coordinates $r,\th,\lambda$. Each observer moves in spacetime with four-velocity $u^\a=c^{-1}dx^\a/d\t$ where $x^\a=\left\{x^0,x^1,x^2,x^3\right\}=\left\{ct,r,\th,\lambda\right\}$ are rotating geodetic coordinates, and $\t$ is the proper time of observer defined in terms of the metric tensor \eqref{27} as follows,
\be\la{ki7}
c^2 d\t^2=-g_{\a\b}(r,\th,\lambda)dx^\a dx^\b\;.
\ee 
Physical space of observers at each instant of time is represented by a three-dimensional hypersurface of constant proper time that is orthogonal everywhere to the world lines of the observers. The metric tensor, $h_{\a\b}$, is given on this hypersurface by \citep{Zelmanov_2006_book,LL_book2}
\be\la{m5vl}
h_{\a\b}\equiv g_{\a\b}+u_\a u_\b\;,
\ee
and is used to measure the spatial distances. Rising and lowering of Greek indices of geometric objects residing on the perturbed manifold are done with the help of the full metric $g_{\a\b}$.

Similarly to classic geodesy, general relativity offers two definitions of relativistic geoid \citep{Soffel_1989_book,Kopejkin_1991} 
\begin{description}
\item[\textcolor{blue}{Definition 1.}] The relativistic $u$-geoid represents a two-dimensional surface at any point of which the rate of the proper time, $\t$, of an ideal clock carried out by a static observer with the fixed coordinates $(r,\th,\lambda)$, is constant. 
\end{description}
The $u$-geoid is determined by equation $W\equiv W(r,\th,\lambda)={\rm const.}$, where the physical gravity potential 
\be\la{18}
W=c^2\left[1-\lt(\frac{d\t}{dt}\ri)\right]_{r,\th,\lambda\;{\rm fixed}}\;.
\ee
It is equivalent to $W/c^2=1-1/u^0$ where $u^0=dt/d\t=(-g_{00})^{-1/2}$ is the time component of the four-velocity of the observer having the fixed coordinates $r,\th,\lambda$, and $g_{00}$ is the time-time component of the metric tensor in the {\it rotating} coordinates. Picking up the value of $u^0$, equation \eqref{18} becomes,
\be\la{19ss}
W(r,\th,\lambda)\equiv c^2\left[1-\left(-g_{00}\ri)^{1/2}\ri]\;.
\ee
This matches the post-Newtonian definition of the $u$-geoid given in previous works \citep{Soffel_1988,Soffel_1989_book,Kopejkin_1991} 
\be\la{pngeoid}
W=W_{\rm N}+\frac1{c^2}W_{\rm pN}+O\left(c^{-4}\right)\;,
\ee
where $W_{\rm N}$ is the Newtonian geoid defined in \eqref{3edg}, and $W_{\rm pN}$ are the post-Newtonian corrections defined in terms of the post-Newtonian potentials entering the post-Newtonian expansion of $g_{00}$ in \eqref{19ss}. We refer the reader to \citep[eq. 18]{Mueller_2008JGeod} and \citep[eq. 8.104]{Kopeikin_2011_book} for further detail.

\begin{description}
\item[\textcolor{blue}{Definition 2.}] The relativistic $a$-geoid represents a two-dimensional surface at any point of which the direction of a plumb line measured by a static observer, is orthogonal to the tangent plane of geoid's surface (\ref{19ss}).

In order to derive equation of $a$-geoid, we notice that the direction of the plumb line is given by a four-vector of the physical acceleration of gravity, $g_\a\equiv -c^2a_\a$ where $a_\a=-\pd_\a\ln u^0$ is a four-acceleration of the static observer given in terms of the time component of its four-velocity (see \eqref{22}). Making use of $W/c^2=1-1/u^0$, we get
\be\la{nu7}
g_\a=-c^2\pd_\a\ln\left(1-\frac{W}{c^2}\ri)\;.
\ee
We consider an arbitrary displacement, $dx^\a_\perp\equiv h^\a{}_\b dx^\b$, on the spatial hypersurface being orthogonal to $u^\a$ everywhere, and make a scalar product of $dx^\a_\perp$ with the direction of the plumb line. It gives,
\be\la{24}
dx^\a_\perp g_\a=dx^\a g_\a=-c^2d\ln\left(1-\frac{W}{c^2}\ri)\;.
\ee
From definition of $a$-geoid the left-hand side of \eqref{24} must vanish due to the condition of orthogonality of the two vectors, $dx^\a_\perp$ and $g_\a$. Therefore, it makes $d\ln\left(1-W/c^2\ri)=0$, which means the constancy of the gravity potential $W$ on the three-dimensional surface of the $a$-geoid. Thus, the surface of $a$-geoid coincides with that of $u$-geoid.   

\end{description}

\section{The anomalous gravity potential}\la{sec4}
\noindent
We define the anomalous (disturbing) gravity potential ${\cal T}\equiv {\cal T}(r,\th,\lambda)$ as the difference between the real gravity potential, $W\equiv W(r,\th,\lambda)$, and the gravity potential, $\bar U(r,\th)$, of the reference matter configuration 
\be\la{28}
{\cal T}(r,\th,\lambda)=W(r,\th,\lambda)-\bar U(r,\th)\;.
\ee
Making use of \eqref{levs1} and \eqref{18} allows us to recast \eqref{28} to
\be\la{28gg}
{\cal T}(r,\th,\lambda)=c^2\left(\frac{d\bar\t}{dt}-\frac{d\t}{dt}\right)_{r,\th,\lambda\;{\rm fixed}}\;,
\ee
which can be further simplified by noticing that
\be\la{31}
\lt(\frac{d\t}{dt}\ri)^2_{r,\th,\lambda\;{\rm fixed}}=-\bar g_{00}\lt(1+\frac{\varkappa_{00}}{\bar g_{00}}\ri)=\lt(\frac1{\bar u^0}\ri)^2\lt(1-(\bar u^0)^2\varkappa_{00}\ri)=
\lt(1-\bar u^\a\bar u^\b\varkappa_{\a\b}\ri)\lt(\frac{d\bar\t}{dt}\ri)^2_{r,\th,\lambda\;{\rm fixed}}\;,
\ee
because the unperturbed four-velocity, $\bar u^\a$ has only a time component, $\bar u^0\not=0$ in the spherical coordinates under consideration.
Accounting for definition \eqref{levs1}, we get an {\it exact} expression for the anomalous gravity potential in the form,
\be\la{32}
{\cal T}=c^2\left(1-\frac{\bar U}{c^2}\right)\lt(1-\sqrt{1-\bar u^\a\bar u^\b\varkappa_{\a\b}}\ri)\;,
\ee
where the term, $\bar U/c^2$, has the same order of magnitude as the metric pertrubation $\bar u^\a\bar u^\b\varkappa_{\a\b}$. 
For practical applications equation \eqref{32} should be linearised by expanding its right-hand side in the Taylor series, and discarding non-linear terms. It yields
\be\la{33}
{\cal T}=\frac{c^2}{2}\bar u^\a\bar u^\b\varkappa_{\a\b}\;,
\ee
where $\varkappa_{\a\b}$ has been defined in \eqref{27}. We emphasize that $\varkappa_{\a\b}$ is the difference between the actual physical metric, $g_{\a\b}$, and the metric $\bar g_{\a\b}$ of the background manifold which is an exact, axially-symmetric solution of Einstein's equations. Thus, $\varkappa_{\a\b}$ should not be confused with the post-Newtonian expansion of the metric $g_{\a\b}$ around a flat spacetime with the Minkowski metric $\eta_{\a\b}={\rm diag}(-1,1,1,1)$.

Our next task is to derive the differential equation for the anomalous gravity potential ${\cal T}$.

\section{The master equation for the anomalous gravity potential}\la{sec5}
\noindent Let us assume that inside Earth the deviation of the real matter distribution from its unperturbed value is described by the symmetric energy-momentum tensor 
\be\la{34}
c^2\mathfrak{T}^{\a\b}=\mathfrak{e}\,u^\a u^\b+\mathfrak{s}^{\a\b}\;,
\ee 
where $u^\a$ is four-velocity, $\mathfrak{e}$ is the energy density, and $\mathfrak{s}^{\a\b}$ is the symmetric stress tensor of the perturbing matter. The stress tensor includes the isotropic pressure (diagonal components) and shear (off-diagonal components), and is orthogonal to $u^\a$, that is $\mathfrak{s}_{\a\b}u^\a=0$. The energy density of the matter perturbation
\be\la{35}
\mathfrak{e}=\mu\left(c^2+\mathfrak{P}\ri)\;,
\ee
where $\mu$ is the mass density - the same as in \eqref{4qq}, and $\mathfrak{P}$ is the internal (compression) energy of the perturbation. 

For further calculations, a more convenient metric variable is
\be\la{36}
l_{\a\b}\equiv-\varkappa_{\a\b}+\frac12\bar g_{\a\b}\varkappa\;,
\ee
where $\varkappa\equiv\bar g^{\a\b}\varkappa_{\a\b}$. The dynamic field theory of manifold perturbations leads to the following equation for $l_{\a\b}$  \citep{Kopeikin_2013PhRvD,Kopeikin_2014AnPhy},
\be\la{37}
l_{\a\b}{}^{|\m}{}_{|\m}+\bar g_{\a\b}{\cal A}^\m{}_{|\m}-2{\cal A}_{\a|\b}-\bar R^\m{}_\a l_{\b\m}-\bar R^\m{}_\b l_{\a\m}-2\bar R_{\a\m\n\b}l^{\m\n}+2F^{\rm m}_{\a\b}=2\kappa\mathfrak{T}_{\a\b}\;,
\ee
where ${\cal A}^\a\equiv l^{\a\b}{}_{|\b}$ is the gauge vector function, depending on the choice of the coordinates, $\bar R_{\a\m\n\b}$ is the Riemann (curvature) tensor of the background manifold depending on the metric tensor $\bar g_{\a\b}$, its first and second derivatives, $\bar R_{\a\b}=\bar g^{\m\n}\bar R_{\m\a\n\b}$ -- the Ricci tensor, and $F^{\rm m}_{\a\b}$ is the tensorial perturbation of the  background matter induced by the presence of the  perturbation $\mathfrak{T}^{\a\b}$ (see \citep[eqs. 148-150]{Kopeikin_2013PhRvD} for particular details). 

In what follows, we focus on derivation of the master equation for the anomalous gravity potential ${\cal T}$ in the exterior space that is outside of  the background matter of the reference configuration. Derivation of the master equation for ${\cal T}$ inside matter will be given somewhere else. To achieve our goal, we introduce two auxiliary scalars, 
\ba\la{38a} 
\mathfrak{q}&\equiv&\bar u^\a\bar u^\b l_{\a\b}+\frac{l}2\;,\\\la{38bb}
\mathfrak{p}&\equiv&\bar h^{\a\b}l_{\a\b}\;,
\ea
where 
\be\la{38b}
l\equiv\bar g^{a\b}l_{\a\b}=2(\mathfrak{p}-\mathfrak{q})\;.
\ee
In terms of the scalar $\mathfrak{q}$ the anomalous gravity potential \eqref{33} reads
\be\la{39}
{\cal T}=-\frac{c^2}2\mathfrak{q}\;,
\ee
where we have used the property $\varkappa=l$.
Taking from both sides of \eqref{39} the covariant Laplace operator yields
\be\la{40}
\Box {\cal T}\equiv -\frac{c^2}2\Box\mathfrak{q}\;,
\ee
where $\Box\mathfrak{q}\equiv\mathfrak{q}^{|\mu}{}_{|\mu}$ is to be calculated from \eqref{37}. 

We notice that according to \citep[eqs. 148-150]{Kopeikin_2013PhRvD} $F^{\rm m}_{\a\b}$ is directly proportional to the thermodynamic quantities of the background matter and, thus, vanishes in the exterior (with respect to the background matter) space. Hence, we can drop off $F^{\rm m}_{\a\b}$ in (\ref{37}) in the exterior-to-matter domain. After contracting \eqref{37} with $\bar g^{\a\b}$, and accounting for \eqref{38b} we obtain 
\be\la{41}
\Box\mathfrak{q}-\mathfrak{p}^{|\a}{}_{|\a}-{\cal A}^\a{}_{|\a}=-\kappa\mathfrak{T}\;,
\ee
where all terms depending on the Ricci tensor $\bar R_{a\b}$ cancel out, $\mathfrak{T}\equiv\bar g^{\a\b}\mathfrak{T}_{\a\b}$, and we still have terms with the gauge field ${\cal A}^\a$. Now, we use the gauge freedom of general relativity to simplify (\ref{41}). More specifically, we impose the gauge condition
\be\la{42}
{\cal A}^\a=-\mathfrak{p}^{\a}\;,
\ee
where $\mathfrak{p}^\a\equiv\bar g^{\a\b}\pd_\b\mathfrak{p}$.
This gauge allows us to eliminate function $\mathfrak{p}$ from \eqref{41} and, after making use of \eqref{40}, reduce it 
to a simple form of a covariant Poisson equation
\be\la{43}
\Box {\cal T}=\frac12c^2\kappa\mathfrak{T}\;.
\ee
In the Newtonian approximation the trace of the energy-momentum tensor is reduced to the negative value of the matter density of the perturbation, $\mathfrak{T}\simeq -\mu$, while $\Box {\cal T}\simeq\Delta_{\rm N}{\cal T}$. Hence,  equation \eqref{43} matches its Newtonian counterpart \eqref{5}. Outside the mass distribution the master equation for the anomalous gravity potential is reduced to the covariant Laplace equation
\be\la{44}
\Box {\cal T}=0\;.
\ee

Equations \eqref{43}, \eqref{44} extend similar equations \eqref{5}, \eqref{6} of classic geodesy to the realm of general relativity. The main difference is that the covariant Laplace operator in \eqref{43}, \eqref{44} is taken in curved space with the metric $\bar g_{\a\b}$. The explicit form of the covariant Laplace operator applied to a scalar ${\cal T}$ in spherical coordinates, $x^i=\{x^1,x^2,x^3\}=\{r,\th, \lambda\}$, reads \citep[Problem 7.7.]{LPPT_1975_book}
\be\la{44dd}
\Box{\cal T}\equiv\frac1{\sqrt{-\bar g}}\pd_i\left(\sqrt{-\bar g}\bar g^{ij}\pd_j{\cal T}\ri)\;,
\ee
where the repeated indices mean the Einstein summation, $\pd_i\equiv \pd/\pd x^i$ is the partial derivative, $\bar g={\rm det}[\bar g_{\a\b}]=-A^4B^2N^2r^4\sin^2\th$ is the determinant of the background metric, and we have assumed that the perturbation is stationary which eliminates all time derivatives in \eqref{44dd}.
It brings (\ref{44}) to the following form
\be\la{45}
\frac{\pd}{\pd r}\lt[BNr^2 \frac{\pd {\cal T}}{\pd r}\ri]+\frac1{\sin\th}\frac{\pd}{\pd \th}\lt[BN\sin\th \frac{\pd {\cal T}}{\pd \th}\ri]+\frac{A^2}{BN\sin^2\th}\lt[N^2-\frac1{c^2}(\O-\mathfrak{G})^2B^2r^2\sin^2\th\ri]\frac{\pd^2 {\cal T}}{\pd\lambda^2}=0\;,
\ee
where functions $A=A(r,\th)$, $B=B(r,\th)$, $N=N(r,\th)$, $\mathfrak{G}=\mathfrak{G}(r,\th)$ are solutions of Einstein's equations \eqref{13}-\eqref{16} for the reference level configuration. 

Equation \eqref{45} can be further simplified by making use of spherically-symmetric approximation and noticing that in this approximation functions $A=B$ \citep{Bardeen_1971ApJ}. Moreover we have \citep{Gourgoulhon_2010arXiv}
\be\la{niurtcv6}
A=(1-\frac{GM}{2c^2r})^2\qquad,\qquad N=\frac{\displaystyle 1-\frac{GM}{2c^2r}}{\displaystyle 1+\frac{GM}{2c^2r}}\;.
\ee  
Neglecting the post-post-Newtonian corrections of the order $1/c^4$ and the gravitomagnetic contribution, we get the post-Newtonian version of equation \eqref{45} in the spherical approximation that reads
\be\la{46ffh}
\D_{\rm N}{\cal T}=\frac{\O^2}{c^2}\frac{\pd^2 {\cal T}}{\pd\lambda^2}\;,
\ee
where the Laplacian $\D_{\rm N}$ has been defined in \eqref{1aa}. The post-Newtonian equation \eqref{46ffh} can be solved by iterations by expanding the distrubing potential in the post-Newtonian series
\be\la{pnT}
{\cal T}=T_{\rm N}+\frac1{c^2}T_{\rm pN}+O(c^{-4})\;,
\ee
where $T_N$ is the Newtonian disturbing potential obeying equation \eqref{6}, and $T_{pN}$ is the post-Newtonian correction.

\section{Geoid's height}\la{sec6}
\noindent
We introduce the relativistic geoid height, ${\cal N}$, by making use of relativistic generalization of 
Bruns' formula (\ref{8}). Let a point ${\cal Q}$ lie on an equipotential reference surface ${\cal S}_1$ and has coordinates $x^\a_{\cal Q}$, and a point ${\cal P}$ lie on another equipotential surface ${\cal S}_2$, and has coordinates $x^\a_{\cal P}$. The height difference, ${\cal N}$, between the two surfaces is defined as the absolute value of the integral taken along the direction of the plumb line passing through the points ${\cal Q}$ and ${\cal P}$,
\be\la{50}
{\cal N}=\int_{\cal Q}^{\cal P}n_\a\frac{dx^\a}{d\ell}d\ell\;,
\ee 
where $n_\a\equiv g_\a/g$ is the unit (co)vector along the plumb line, $g_\a$ is the relativistic acceleration of gravity \eqref{nu7}, $g\equiv(h^{a\b}g_\a g_\b)^{1/2}$, and $\ell$ is the proper length defined in space by \citep{Zelmanov_2006_book,LL_book2} 
\be\la{51}
d\ell^2=\bar h_{\a\b}dx^\a dx^\b\;.
\ee
In case, when the height difference is small enough, we can use the second mean value theorem for integration \citep{Hobson_1909} and approximate the integral in (\ref{50}) as follows
\be\la{52}
{\cal N}=\int_{\cal Q}^{\cal P}\frac{g_\a(x) dx^\a}{g(x)}=-\frac{c^2}{g_{\cal Q}}\int_{\cal Q}^{\cal P}\pd_\a\ln\left(1-\frac{W}{c^2}\right)dx^\a=\frac{c^2}{g_{\cal Q}}\ln\left|\frac{\displaystyle 1-\frac{W({\cal Q})}{c^2}}{\displaystyle 1-\frac{W({\cal P})}{c^2}}\right|\;,
\ee
where $g_{\cal Q}=g({\cal Q})$ denotes the magnitude of the relativistic acceleration of gravity taken on the equipotential surface ${\cal S}_1$.
Equation \eqref{52} is exact. Separation of the height ${\cal N}$ in the Newtonian part and the post-Newtonian corrections depends on how we define the reference equipotential surface ${\cal S}_1$.  

Let us choose the reference surface by equation $W({\cal Q})=\bar U$ where $\bar U$ is the exact solution of the Einstein equations described in Section \ref{sec2aa}. Then, expanding the logarithm in \eqref{52} with respect to the ratio $W/c^2$ and making use of definition \eqref{28} of the anomalous gravity potential ${\cal T}$, we obtain from (\ref{52})
\be\la{46}
{\cal N}=\frac{|{\cal T}(\cal P)|}{\g_{\cal Q}}\;,
\ee
where the disturbing potential, ${\cal T}$, is measured at the point ${\cal P}$ on the geoid surface $W$, and the acceleration of gravity $\g_{\cal Q}\equiv g_{\cal Q}$ is measured at point ${\cal Q}$ on the reference surface $\bar U$. 

Relativistic Bruns' formula (\ref{46}) yields geoid's undulation with respect to the unperturbed reference level surface in general relativity. Because we have defined this surface as an equipotential surface $\bar U$ of the exact (unperturbed) solution of the Einstein equations, the height ${\cal N}$ does not represent the undulation of the relativistic geoid $W$ with respect to the Newtonian equipotential surface $\bar U_N$ defined by equation \eqref{1}. Expansion of the height ${\cal N}$ in \eqref{46} in the post-Newtonian series around the value of the surface $\bar U$, yields 
\be\la{62}
{\cal N}={\mathfrak N}+\frac1{c^2}{\mathfrak N}_{\rm pN}+O\left(c^{-4}\right)\;,
\ee 
where ${\mathfrak N}$ is the classic definition \eqref{8} of the geoid height given in terms of the Newtonian disturbing potential \eqref{4}. The post-Newtonian correction ${\mathfrak N}_{\rm pN}$ to the height ${\mathfrak N}$ has a magnitude of the order ${\mathfrak N}_{\rm pN}\simeq(V_{\rm N}/c^2)\times{\mathfrak N}$, where $V_{\rm N}$ is the Newtonian gravitational potential of the Earth. Because the largest undulation of the Newtonian geoid of the Earth does not exceed 100 meters \citep{Torge_2012_book}, the post-Newtonian correction to the undulation is exceedingly small, ${\mathfrak N}_{\rm pN}\simeq 7\times 10^{-6}$ cm. Exact equation for ${\mathfrak N}_{\rm pN}$ will be published somewhere else.

On the other hand, we can choose the reference surface ${\cal S}_1$ coinciding with the Newtonian equipotential surface $W({\cal Q})=\bar U_{\rm N}$, where $\bar U_{\rm N}$ is defined in \eqref{1}. Then, expanding $W({\cal P})$ in the post-Newtonian series \eqref{pngeoid} and taking into account definition \eqref{4} of the Newtonian disturbing potential $T_{\rm N}$, we obtain from \eqref{52}
\be\la{69}
{\cal N}={\mathfrak N}+\frac1{c^2}\frac{W_{\rm pN}(\cal P)}{\g_{\cal Q}}+O\left(c^{-4}\right)\;,
\ee
where the second term in the right-hand side defines the relativistic correction to the geoid undulation with respect to the Newtonian reference ellipsoid. Formula \eqref{69} coincides with the expression for the post-Newtonian undulation of the relativistic geoid given in \citep[eq. 19]{Mueller_2008JGeod}, and it amounts to a few millimetres. 

Notice that due to the two different possible choices of the reference level surface, ${\cal S}_1$, the post-Newtonian height's correction, ${\mathfrak N}_{\rm pN}\ll W_{\rm pN}/\g_{\cal Q}$. It remains up to geodesists to decide what definition of the reference surface and the geoid height is the most meaningful in practical applications. The options are:
\begin{enumerate} 
\item calculate the post-Newtonian reference level surface $\bar U$ by solving the Einstein equations and operate with the Newtonian-like Bruns' formula to determine geoid's undulation with respect to $\bar U$, 
\item operate with the Newtonian reference level surface $\bar U_{\rm N}$ and calculate geoid's undulation from the post-Newtonian version of Bruns' formula \eqref{69} where $W_{\rm pN}$ is found by solving the Einstein equations.
\end{enumerate} 
The first choice seems to us be more preferable because the second case requires re-calculation of the Newtonian gravity potential $V_{\rm N}$ to take into account the post-Newtonian corrections to geoid's figure. Discussion of this problem will be continued somewhere else.

\section*{Acknowledgement}   
We are grateful to two anonymous referees for numerous critical remarks and valuable suggestions which helped us to significantly improve the manuscript. One of us (S.M.K.) thanks Dr. D. Petroff for fruitful conversations on the relativistic theory of rotating fluids and the post-Newtonian figures of equilibrium and for providing us with friendly and exhaustive explanation of the subtle theoretical details of his work \citep{Petroff_2003PhRvD}. 

The present work has been supported by the Faculty Fellowship 2014 in the College of Arts and Science of the University of Missouri and the grant 14-27-00068 of the Russian Scientific Foundation.

\bibliographystyle{elsarticle-num}
\bibliography{relativistic_geodesy}
\end{document}